\documentclass[conference,a4paper]{IEEEtran}

\usepackage[english]{babel}
\usepackage{amsmath,amssymb,amsfonts,mathrsfs}
\usepackage[amsmath,thmmarks]{ntheorem}
\usepackage{graphicx}
\usepackage{tikz}
\usetikzlibrary {positioning}
\usetikzlibrary{decorations.text}
\usepackage{subfigure}

\newtheorem{protocol}{Protocol}
\newtheorem{theorem}{Theorem}
\newtheorem{lemma}{Lemma}

\begin{document}

\sloppy

\title{Trading Permutation Invariance for Communication in Multi-Party Non-Locality Distillation}

 \author{
   \IEEEauthorblockN{Helen Ebbe and Stefan Wolf}
   \IEEEauthorblockA{Faculty of Informatics\\
     University of Lugano (USI)\\
     6900 Lugano, Switzerland\\
     Email: \{ebbeh,wolfs\}@usi.ch} 
 }
\maketitle

\begin{abstract}
Quantum theory puts forward  phenomena unexplainable by classical
physics---or information, for that matter. A prominent example is
\emph{non-locality}. Non-local correlations  cannot be
explained, in classical terms, by shared information  but only by communication. On the other hand, 
the phenomenon does not \emph{allow} for  (potentially
faster-than-light) message transmission. 
The fact that some non-local and non-signaling correlations are predicted by quantum
theory, whereas others fail to be, asks for a
criterion, as simple as possible, that characterizes which
joint input-output behaviors  are ``quantum'' and which are not. In the context of the
derivation  of such criteria, it is of central importance to
understand when non-local correlations can be {\em amplified\/}
by a non-interactive protocol, {\em i.e.}, whether some types 
of weak non-locality can be distilled into stronger by local 
operations. Since it has been recognized that the searched-for criteria 
must inherently be {\em multi-partite}, the question of distillation,
extensively studied and understood {\em two\/}-party scenarios,
should be adressed in the multi-user setting, where much less is known. 
Considering the space of intrinsically $n$-partite correlations, we show the
possibility of distilling weak non-local boxes  to the algebraically
maximal ones without any communication. Our protocols improve 
on previously known methods which  still required partial communication. The price
we have to pay for dropping the need for communication entirely
is the assumption of {\em permutation invariance\/}: Any correlation that 
can be realized between \emph{some} set of players is possible between {\em any\/}
such set. This assumption is very natural since the laws of physics
are invariant under spacial translation.
\end{abstract}

\section{Motivation and Outline}
Einstein, Podolsky, and Rosen \cite{EPR35} raised the question ``Can
quantum-mechanical description of physical reality be considered
complete?'' In direct response to that question, Bell \cite{B64}
showed that quantum mechanics is incompatible with a \emph{local
  hidden variable theory\/}: The theory predicts correlations that
are, 
in classical terms, not explainable by shared information but only 
by communication. It is important to note, \emph{however}, that, on the 
other hand, the arising correlations do not allow for message transmission.

With the goal of a systematic and generalized understanding of
non-local correlations, quantum and beyond, 
Popescu and Rohrlich described an input-output behavior that maximally
violates the \emph{Bell-inequality}, but that still fulfills the
non-signaling condition~\cite{PR94}. Their bipartite input-output
behavior or {\em box\/} can classically be realized with 75\% only, by
quantum states with 85\%~\cite{C80}, 
whereas even the perfect approximation would still be compatible with 
the non-signaling principle. This means that, strangely enough,
quantum physics is not maximally non-local and cannot be singled out
by the non-signaling principle. It is a fascinating and conceptually
important question whether there is an (information-theoretical)
principle that is able to describe exactly the quantum
correlations. One such attempt has been to generalize the
non-signaling principle to parties that are allowed to use limited
communication, to the so-called \emph{information-causality
  principle}~\cite{P09}. 
Other authors have looked for principles characterizing
quantum correlations as the ones that do not improve the efficiency of \emph{non-local
  computation}~\cite{LPSW07} or do not collapse
\emph{communication complexity} \cite{BBLMTU06}.
 Two more physically motivated principles are  \emph{macroscopic
   locality} \cite{NW10} and \emph{local orthogonality}~\cite{FSA12}. 
In each case, it has been shown that quantum physics respects the 
corresponding principle, whereas some ``super-quantum'' correlations 
violate it. For none of the principles, however, it was possible to
show that {\em every\/} non-quantum behavior is in violation.

In the search of a principle {\em exactly\/} singling out
quantum
theory,  the possibility of making (weak) non-local correlations
stronger by local wirings is paramount since it offers the possibility of generating systems violating some
principle from correlations which respect it. 
Therefore,
a systematic understanding of the power and limitations of
distillation of non-locality potentially leads to deep insights into
the mysterious nature of quantum theory.

The question of non-locality distillation was mainly studied in the
bipartite scenario: There exist weak non-local correlations that can
be distilled to an almost perfect Popescu-Rohrlich box by an adaptive
protocol~\cite{FWW09,BS09,HR10}. On the other hand, {\em isotropic\/}
correlations seem to be undistillable~\cite{DW08}.

In the context both of information principles able to single out 
quantum theory \cite{mult11} as well as for information-processing 
tasks such as {\em randomness generation\/} \cite{rand12}, it has turned out that {\em
  multi-party},
as opposed to only bipartite correlations, play a crucial role.
Nevertheless, much less is known for that case.
One effort was to generalize
the XOR protocol~\cite{HW10}, but it fails to distill maximal
non-local boxes. It was shown in \cite{EW13a,EW13b} that the large
class of~\emph{full-correlation boxes} can be distilled by a
multipartite version of Brunner-Skrzypczyk's protocol~\cite{BS09}
under the (strong) assumption that {\em partial communication\/}
is allowed. This latter assumption, unfortunately, puts into question 
the relevance of the protocol in the context described above, namely of 
finding information-based criteria singling out quantum theory.

We introduce a new kind of multi-party distillation protocols by
showing that the need for communication can be dropped entirely.
The price for this is the need for the assumption that any correlation
which can be realized between \emph{some} set of players is also possible 
between any \emph{other} set. We believe this assumption to be quite natural
since we imagine the correlations to arise from the interaction with 
a concrete physical system. In this sense, the assumption is true in 
every world in which the laws of physics are invariant under spacial 
translation. 

This is an outline of the present article. 
We first characterize full-correlation boxes and give for them a
criterion for being maximally non-local (Section
\ref{sec:full}). Second, we present a new distillation protocol that
does not use partial communication, but that is still able to distill
every non-isotropic faulty version of a full-correlation box to a
close-to-perfect one. This new protocol requires  the different
parties to be able to arbitrarily distribute the input-and output-interfaces of the weak boxes  (Section~\ref{sec:dist}). In Section \ref{sec:ex}, we illustrate our result with an example.

\section{Definitions}
\subsection{Systems and Boxes}
In an \emph{$n$-partite input-output system}, each of the $n$ parties inputs an element $x_i$ and receives immediately an output~$a_i$. The behavior of this system is defined by a conditional distribution
\begin{equation}
P_{A_1A_2\cdots A_n\vert X_1X_2\cdots X_n}\ ,
\end{equation}
where $X_i$ is the input and $A_i$ the output variable of the $i$th party.

An $n$-partite system with conditional probability distribution $P\left( a_1 a_2 \cdots a_n \vert x_1 x_2 \cdots x_n\right)$ is said \emph{non-signaling} if the marginal distribution for each subset of parties $\lbrace a_{k_1} , a_{k_2},..., a_{k_m}\rbrace$ only depends on its their own inputs
\begin{equation}
P\left( a_{k_1} \cdots a_{k_m}\vert x_1 \cdots x_n\right)  = P\left( a_{k_1} \cdots a_{k_m}\vert x_{k_1}  \cdots x_{k_m}\right)\ .
\end{equation}

If the $n$-partite system fulfills the \emph{non-signaling condition}, \emph{i.e.,} the system cannot be used to transmit instantaneously information from one party to another one, then this system is called \emph{n-partite box}.

\subsection{Multipartite Locality}
Of central interest for us are $n$-partite boxes with the property that the parties cannot simulate the behavior of the box without communication, but by shared randomness only. This property is called \emph{non-locality}.

An $n$-partite box with input variables $X_1$, $X_2$, ..., $X_n$ and output variables $A_1$, $A_2$, ..., $A_n$ is \emph{local} if
\begin{equation}
P_{A_1 A_2 \cdots A_n|X_1 X_2 \cdots X_n}=\sum_{r\in{\cal R}}{P_R(r)\cdot P_{A_1|X_1}^r\cdots P_{A_n|X_n}^r}\
\end{equation}
for some random variable $R$.

\section{Full-Correlation Boxes}
\label{sec:full}
\subsection{Definition and Related Boxes}
We focus our attention to one of the most general type of boxes: the \emph{$n$-partite full-correlation boxes} that were introduced by Barrett and Pironio \cite{BP05}. This kind of boxes has the property that it displays  correlation only with respect to the \emph{full} set of parties.
An \emph{$n$-partite full-correlation box} takes as inputs $\textbf{x}=(x_1,x_2, \dots x_n)$ and as outputs $\textbf{a}=(a_1,a_2, \dots a_n)$, where all $x_i, a_i \in \{0, 1\}$. The input-output behavior is characterized by the following conditional distribution:
\begin{equation}
P(\textit{\textbf{a}}\vert \textit{\textbf{x}}) = \begin{cases} \frac{1}{2^{n-1}}&\text{$\sum\limits_i a_i \equiv f(\textit{\textbf{x}})$ (mod 2)}\\0&\text{otherwise,}\end{cases}
\end{equation}
where $f(\textit{\textbf{x}})$ is a Boolean function of the inputs. 

A special case of the full-correlation boxes is the $n$-party generalization of the \emph{Popescu-Rohrlich box} \cite{PR94} ($n$-PR box) that is defined by the conditional distribution:
\begin{equation}
P^{\text{PR}}_n(\textit{\textbf{a}}\vert \textit{\textbf{x}}) = \begin{cases} \frac{1}{2^{n-1}}&\bigoplus\limits_{i=1}^n a_i = \prod\limits_{i=1}^n x_i\\0&\text{otherwise.}\end{cases}
\end{equation}

If the output does not depend on all inputs, then we call the box an $(n,k)$-PR box ($k\leq n$):
\begin{equation}
P^{\text{PR}}_{(n,k)}(\textit{\textbf{a}}\vert \textit{\textbf{x}}) = \begin{cases} \frac{1}{2^{n-1}}&\bigoplus\limits_{i=1}^n a_i = \prod\limits_{i=1}^k x_i\\0&\text{otherwise.}\end{cases}
\end{equation}

Note that an $n$-PR box and an $(n,n)$-PR box are identical.

\subsection{Construction of $n$-Partite PR Boxes}
\label{sec:constr}
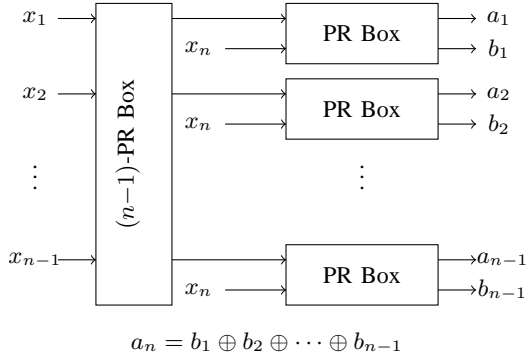
\begin{figure}[!htb]
\begin{center}
\begin{tikzpicture}
\begin{small}
\node [draw, rectangle, minimum width=1cm, minimum height=4cm] (A) at (0,0) {};
\node [draw, rectangle, minimum width=2cm, minimum height=0.8cm] (B)  at (3,1.6) {PR Box};
\node [draw, rectangle, minimum width=2cm, minimum height=0.8cm] (C) at (3,0.6) {PR Box};
\node [draw, rectangle, minimum width=2cm, minimum height=0.8cm] (D) at (3,-1.6) {PR Box};
\draw [decoration={text along path,
       text={$(n-1)$-PR Box},text align={center},text color=black},decorate]  (0,-1) to []  (0,1);

\node [] () at (-1.3,-0.2) {$\vdots$};
\node [] () at (3,-0.2) {$\vdots$};

\draw[->] (0.5,1.8) to[] (2,1.8);
\draw[->] (0.5,0.8) to[] (2,0.8);
\draw[->] (0.5,-1.4) to[] (2,-1.4);
\draw[->] (1.2,1.4) to[] (2,1.4);
\draw[->] (1.2,0.4) to[] (2,0.4);
\draw[->] (1.2,-1.8) to[] (2,-1.8);
\node [] () at (0.85,1.4) {$x_n$};
\node [] () at (0.85,0.4) {$x_n$};
\node [] () at (0.85,-1.8) {$x_n$};

\draw[->] (-1,1.8) to[] (-0.5,1.8);
\draw[->] (-1,0.8) to[] (-0.5,0.8);
\draw[->] (-1,-1.4) to[] (-0.5,-1.4);
\node [] () at (-1.3,1.8) {$x_1$};
\node [] () at (-1.3,0.8) {$x_2$};
\node [] () at (-1.3,-1.4) {$x_{n-1}$};

\draw[->] (4,1.8) to[] (4.5,1.8);
\draw[->] (4,0.8) to[] (4.5,0.8);
\draw[->] (4,-1.4) to[] (4.5,-1.4);
\draw[->] (4,1.4) to[] (4.5,1.4);
\draw[->] (4,0.4) to[] (4.5,0.4);
\draw[->] (4,-1.8) to[] (4.5,-1.8);
\node [] () at (4.8,1.8) {$a_1$};
\node [] () at (4.8,0.8) {$a_2$};
\node [] () at (4.85,-1.4) {$a_{n-1}$};
\node [] () at (4.8,1.4) {$b_1$};
\node [] () at (4.8,0.4) {$b_2$};
\node [] () at (4.85,-1.8) {$b_{n-1}$};

\node [] () at (1.75,-2.5) {$a_n = b_1 \oplus b_2 \oplus \dots \oplus b_{n-1}$};

\end{small}
\end{tikzpicture}
\end{center}
\caption[Recursive Construction of an $n$-PR Box]{Recursive Construction of an $n$-PR Box}
\label{fig:nPR}
\end{figure}

In \cite{BLMPPR05} is shown how a $3$-PR box can be constructed from three PR boxes. This construction can be used to construct an $n$-PR recursively: Assume that the first $n-1$ parties share a $n-1$-PR box and every of this parties input their input $x_i$ to this box. The outputs (say $a_i'$) of the box fulfills \begin{equation}
{i=1}^{n-1} a_i' =\bigwedge\limits_{i=1}^{n-1} x_i\ .
\end{equation}
Every of these $n-1$ parties inputs the output bit in the PR box shared with the $n$th party and the $n$th party inputs in every PR box his input bit. The output bit of the $n$th party is the XOR of all his outputs from the PR boxes (see Fig. \ref{fig:nPR}). In the end, the outputs fulfill 
\begin{eqnarray}
\bigoplus\limits_{i=1}^{n} a_i & = &\bigoplus\limits_{i=1}^{n-1} a_i\oplus \bigoplus\limits_{i=1}^{n-1} a_i \nonumber \\
& = & \bigoplus\limits_{i=1}^{n-1} (a_i\oplus b_i) \nonumber \\
& = & \bigoplus\limits_{i=1}^{n-1} (x_n\wedge a_i') \nonumber \\
& = & \bigwedge\limits_{i=1}^{n} x_i \ .
\end{eqnarray}

\subsection{Construction of Full-Correlation Boxes}
In the same way as in \cite{EW13a, EW13b}, we look how full-correlation boxes can be constructed by generalized PR-boxes and how they can be characterized.

\begin{lemma}
\label{lem:boolf}
If $f$ is a Boolean function of the input elements $x_1,x_2,...,x_n$, then it can be written as
\begin{equation}
f(x_1,...,x_n) = \bigoplus\limits_{I \in \mathcal{I}} \left( a_I\cdot \bigwedge\limits_{i \in I}x_i \right)\ ,
\end{equation}
where $\mathcal{I} = \mathcal{P}\left( \lbrace 1,2,...,n \rbrace\right) $ and $a_I \in \lbrace 0, 1\rbrace$ for all $I \in \mathcal{I}$.
\end{lemma}

\begin{figure}[!htb]
\begin{center}
\begin{tikzpicture}
\begin{small}
\begin{scriptsize}
\node [draw, rectangle, minimum width=5cm, minimum height=1.5cm] (A) at (0,0) {};
\node [draw, rectangle, minimum width=1.3cm, minimum height=0.4cm] (B)  at (0,0) {3-PR Box};
\node [draw, rectangle, minimum width=1.3cm, minimum height=0.4cm] (C) at (1.575,0) {3-PR Box};
\node [draw, rectangle, minimum width=1.3cm, minimum height=0.4cm] (D) at (-1.575,0) {3-PR Box};
\end{scriptsize}
\node [draw, rectangle, minimum width=2cm, minimum height=1cm] (E) at (-5.5,0) {$1\oplus xy\oplus xz$};
\draw[->, thick] (-3,0) to[] (-4,0);
\draw[->, thick] (-4,0) to[] (-3,0);

\draw[->, thick] (-5.5,0.75) to[] (-5.5,0.5);
\draw[->, thick] (-6.1,0.75) to[] (-6.1,0.5);
\draw[->, thick] (-4.9,0.75) to[] (-4.9,0.5);
\node [] () at (-5.5,1) {$y$};
\node [] () at (-6.1,1) {$x$};
\node [] () at (-4.9,1) {$z$};

\draw[->, thick] (-5.5,-0.5) to[] (-5.5,-0.75);
\draw[->, thick] (-6.1,-0.5) to[] (-6.1,-0.75);
\draw[->, thick] (-4.9,-0.5) to[] (-4.9,-0.75);
\node [] () at (-5.5,-1) {$b$};
\node [] () at (-6.1,-1) {$a$};
\node [] () at (-4.9,-1) {$c$};

\draw[->, thick] (0,1) to[] (0,0.75);
\draw[->, thick] (1.575,1) to[] (1.575,0.75);
\draw[->, thick] (-1.575,1) to[] (-1.575,0.75);
\node [] () at (0,1.25) {$y$};
\node [] () at (1.575,1.25) {$z$};
\node [] () at (-1.575,1.25) {$x$};

\draw[->, thick] (0,-0.75) to[] (0,-1.3);
\draw[->, thick] (1.575,-0.75) to[] (1.575,-1);
\draw[->, thick] (-1.575,-0.75) to[] (-1.575,-1);
\node [] () at (0,-1.7) {$b = b_1\oplus b_2\oplus b_3$};
\node [] () at (1.575,-1.25) {$c = c_1\oplus c_2\oplus c_3$};
\node [] () at (-1.575,-1.25) {$a = a_1\oplus a_2\oplus a_3$};

\draw[->, thick] (0,0.4) to[] (0,0.2);
\draw[->, thick] (1.575,0.4) to[] (1.575,0.2);
\draw[->, thick] (-1.575,0.4) to[] (-1.575,0.2);
\node [] () at (0,0.55) {$y$};
\node [] () at (1.575,0.55) {$1$};
\node [] () at (-1.575,0.55) {$1$};

\draw[->, thick] (0,-0.2) to[] (0,-0.4);
\draw[->, thick] (1.575,-0.2) to[] (1.575,-0.4);
\draw[->, thick] (-1.575,-0.2) to[] (-1.575,-0.4);
\node [] () at (0,-0.55) {$b_2$};
\node [] () at (1.575,-0.55) {$b_3$};
\node [] () at (-1.575,-0.55) {$b_1$};

\draw[->, thick] (0.5,0.4) to[] (0.5,0.2);
\draw[->, thick] (2.075,0.4) to[] (2.075,0.2);
\draw[->, thick] (-1.075,0.4) to[] (-1.075,0.2);
\node [] () at (0.5,0.55) {$1$};
\node [] () at (2.075,0.55) {$z$};
\node [] () at (-1.075,0.55) {$1$};

\draw[->, thick] (0.5,-0.2) to[] (0.5,-0.4);
\draw[->, thick] (2.075,-0.2) to[] (2.075,-0.4);
\draw[->, thick] (-1.075,-0.2) to[] (-1.075,-0.4);
\node [] () at (0.5,-0.55) {$c_2$};
\node [] () at (2.075,-0.55) {$c_3$};
\node [] () at (-1.075,-0.55) {$c_1$};

\draw[->, thick] (-0.5,0.4) to[] (-0.5,0.2);
\draw[->, thick] (1.075,0.4) to[] (1.075,0.2);
\draw[->, thick] (-2.075,0.4) to[] (-2.075,0.2);
\node [] () at (-0.5,0.55) {$x$};
\node [] () at (1.075,0.55) {$x$};
\node [] () at (-2.075,0.55) {$1$};

\draw[->, thick] (-0.5,-0.2) to[] (-0.5,-0.4);
\draw[->, thick] (1.075,-0.2) to[] (1.075,-0.4);
\draw[->, thick] (-2.075,-0.2) to[] (-2.075,-0.4);
\node [] () at (-0.5,-0.55) {$a_2$};
\node [] () at (1.075,-0.55) {$a_3$};
\node [] () at (-2.075,-0.55) {$a_1$};

\end{small}
\end{tikzpicture}
\end{center}
\caption[Construction of the $1\oplus xy \oplus xz$-Box]{Construction of the $1\oplus xy \oplus xz$-Box}
\label{fig:equivalenz}
\end{figure}
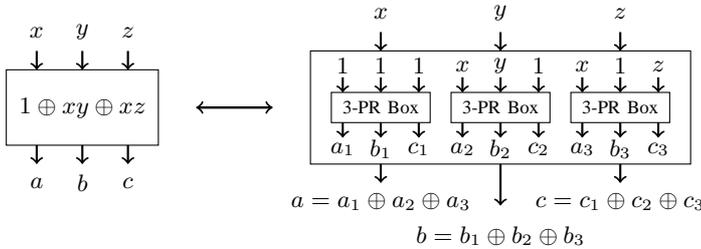

Hence, it is obvious that the full-correlation box associated to the Boolean function $f$ can be constructed by $\sum_{I \in \mathcal{I}}a_I$ $n$~-~PR boxes. For an example, see Fig.~\ref{fig:equivalenz}. Note that the $n$-PR boxes belonging to an $a_I$ where $\vert I \vert \leq 1$ are local and can be simulated by local operations and shared randomness.

We define the set of all non-local $n$-PR boxes that are needed to simulate the full-correlation box: Let
\begin{equation}\label{equ:j}
\mathcal{J} := \lbrace I \in \mathcal{I}\, \vert\, a_I = 1 \text{ and } \vert I\vert \geq 2\rbrace\ .
\end{equation}
This set can be partitioned into pairwisely disjoint subsets $\lbrace J_1, J_2, ..., J_{n_\mathcal{J}}\rbrace$ such that all $A \in J_i$ and $B\in J_j$ fulfill $A\cap B = \emptyset$ for all $i \neq j$. We define the maximal number of such subsets as $n_{\mathcal{J}}$ and denote this partition as the \emph{empty-overlap partition of $\mathcal{J}$}.
\subsection{Non-Locality of Full-Correlation Boxes}
We write the Boolean function that characterizes a full-correlation box as in Lemma \ref{lem:boolf}, so it is easy to determine if this box is local or not.

It is obvious that a full-correlation box is local if the associated Boolean function can be written as the XOR of single inputs of the box and a constant. Assume that the function consists of at least one AND-term, then this box can be reduced to a PR box by distributing all input-and output- interfaces only to two parties such that both of them get at least one input that belongs to the AND-term.
Therefore, a full-correlation box is local if and only if the Boolean function can be written as the XOR of a constant and single inputs of the box

\begin{equation}
f(x_1,...,x_n) = \bigoplus\limits_{I \in \mathcal{L}} \left( a_I\cdot \bigwedge\limits_{i \in I}x_i \right)\ ,
\end{equation}
where $\mathcal{L} = \lbrace \emptyset, \lbrace x_1\rbrace, \lbrace x_2\rbrace, ..., \lbrace x_n\rbrace \rbrace$ and $a_I \in \lbrace 0, 1\rbrace$ for all $I \subseteq \mathcal{L}$.

We show that for every non-local full-correlation box, there exists a closest local box (measured in the $L^1$-norm) that is also a full-correlation box.
Let $P$ be the joint probability distribution of a non-local full-correlation box. Then the joint probability distribution of the \emph{closest local full-correlation box $P^*$} is defined by
\begin{equation}
\Vert P - P^* \Vert_1 = \underset{P' \textnormal{ loc. full-corr. box}}{\textnormal{min}}\left( \Vert P - P'\Vert_1\right)\ .
\end{equation}

\begin{lemma}
Let P be the joint probability distribution of an $n$-partite full-correlation box. Then the closest local full-correlation box is one of the closest local boxes. That means

\begin{equation}
\Vert P - P^* \Vert_1  = \textnormal{min}\left( \left\Vert P - \sum_{r\in{\cal R}}{P_R(r)\cdot P_{A_1|X_1}^r\cdots P_{A_n|X_n}^r} \right\Vert_1 \right) \ ,
\end{equation}
for some random variable $\mathcal{R}$.
\end{lemma}

\begin{IEEEproof}
It is obvious that every deterministic local strategy $r \in \mathcal{R}$ can achieve at most the same number of input-output behaviors (XOR of the outputs equal to a Boolean function of the inputs) as the closest local full-correlation box. So every local box (that is a convex combination of these deterministic local strategies) has at least the same distance from the given full-correlation box as the closest local full-correlation box.
\end{IEEEproof}

\subsection{Extremal Boxes of the Non-Signaling Polytope}
It is a well-known fact that all full-correlation boxes are non-signaling, since they can be simulated by PR boxes \cite{BP05}. In \cite{PBS11}, all tripartite extremal boxes of the non-signaling polytope have been characterized, but for more parties it is not known which of the (full-correlation) boxes are extremal.
\begin{theorem}[Extremal Full-Correlation Boxes]
\label{thm_fullcorr}
Let $P$ be an $n$-partite full-correlation box associated to the Boolean function $f$ that depends on $k$ input variables. Then $P$ is an extremal box of the non-signaling polytope if and only if $n_\mathcal{J} = 1$ and $k=n$ hold.
\end{theorem}

Theorem \ref{thm_fullcorr} follows from Lemmas \ref{lemma2}, \ref{lemma3}, \ref{lemma5}, and \ref{lemma6}.

\begin{lemma}
\label{lemma2}
Let $P$ be an $n$-partite full-correlation box with associated function $f$ that depends on $k$ input variables. If $k \neq n$, then $P$ is not an extremal box.
\end{lemma}

\begin{IEEEproof}
$P$ can be written as a convex combination of the following two non-signaling boxes:
\begin{equation}
P^1(\textbf{a}\vert \textbf{x}) = \begin{cases} \frac{1}{2^{k-1}}&\bigoplus\limits_{i=1}^{k}a_i = f(x_1,x_2,...,x_k) \\ &  \text{and $\bigoplus\limits_{i=k+1}^{n}a_i =0$}\\0&\text{otherwise,}\end{cases}
\end{equation}
and
\begin{equation}
P^2(\textbf{a}\vert \textbf{x}) = \begin{cases} \frac{1}{2^{k-1}}&\bigoplus\limits_{i=1}^{k}a_i = 1\oplus f(x_1,x_2,...,x_k) \\ & \text{and $\bigoplus\limits_{i=k+1}^{n}a_i =1$}\\0&\text{otherwise.}\end{cases}
\end{equation}
So $P = \frac{1}{2}P^1 + \frac{1}{2}P^2$.
Therefore, $P$ is not an extremal box of the non-signaling polytope.
\end{IEEEproof}

\begin{lemma}
\label{lemma3}
Let $P$ be an $n$-partite full-correlation box with associated function $f$ that depends on $k$ input variables. Let $k =n$. If $n_\mathcal{J} \geq 2$, then $P$ is not extremal.
\end{lemma}

\begin{IEEEproof}
Since $n_\mathcal{J}$ is at least $2$, we are able to split the Boolean function $f$ in two other Boolean functions, $f_1$ and $f_2$, such that they do not depend on the same input variables. Without loss of generality, we assume that $f_1$ depends on the input variables $x_1, x_2, ..., x_m$ and $f_2$ depends on $x_{m+1}, ..., x_n$ ($m<n$). Therefore, $f$ can be written as $f(x_1,..., x_n) = f_1(x_1, ..., x_m) \oplus f_2(x_{m+1}, ..., x_n)$. So the box $P$ can be written as a convex combination of the following two boxes:
\begin{equation}
P^1(\textbf{a}\vert \textbf{x}) = \begin{cases} \frac{1}{2^{n-2}}&\bigoplus\limits_{i=1}^{m}a_i = f_1(x_1,x_2,...,x_m) \\ & \text{and $\bigoplus\limits_{i=m+1}^{n}a_i = f_2(x_{m+1},..., x_n)$ }\\0&\text{otherwise,}\end{cases}
\end{equation}
and
\begin{equation}
P^2(\textbf{a}\vert \textbf{x}) = \begin{cases} \frac{1}{2^{n-2}}&\bigoplus\limits_{i=1}^{m}a_i = \neg f_1(x_1,x_2,...,x_m) \\ & \text{and $\bigoplus\limits_{i=m+1}^{n}a_i = \neg f_2(x_{m+1},..., x_n)$ }\\0&\text{otherwise.}\end{cases}
\end{equation}
So $P = \frac{1}{2}P^1 + \frac{1}{2}P^2$.
Therefore, $P$ is not an extremal box of the non-signaling polytope.
\end{IEEEproof}

\begin{lemma}[Existence]
\label{lemma5}
Every $n$-PR box is extremal.
\end{lemma}

\begin{IEEEproof}
The proof is based on the same argument as in \cite{HRW10} for showing that \emph{any non-locality implies some secrecy}.
Assume that the $n$-PR box $P$ can be written as a convex combination of two other non-signaling boxes $P^1$ and $P^2$
\begin{equation}
P = \epsilon P^1 + (1 - \epsilon) P^2,
\end{equation} 
where $0 < \epsilon <1$. It is obvious that both of the boxes must fulfill that the XOR of their output elements is equal to the AND of their input elements, \emph{i.e.},
\begin{equation}
\label{equ:prob}
\text{Prob}\left[ \bigoplus\limits_{i=1}^{n} A_i = \prod\limits_{i=1}^{n} X_i\mid X_i = x_i ~\forall ~1\leq i\leq n \right] = 1
\end{equation}
for all input elements $x_i \in \{0, 1\}$. We will show that all possible biases, $p_i := \text{Prob}\left[ A_i=0 \vert X_k=0 \text{ for all } k\right] $ for all $1\leq i \leq n-1$ such that the box is non-signaling, must be $p_i = 1/2$. Therefore, $P$ cannot be written as a convex combination of other non-signaling boxes.

Assume without loss of generality that all $p_i \geq 1/2$ for all $1\leq i \leq n-1$. Because of Equation (\ref{equ:prob}), the bias $p_n$ can be computed from  the biases $p_i$ for $i \in \lbrace 1, 2, ..., n-1\rbrace$.

Since our box is non-signaling, all biases are independent of the other parties' inputs. We determine step by step the biases $p'_i := \text{Prob}\left[ A_i=0 \vert X_i=1\right]$ for all $i$ and get that $p'_i = p_i$. If not all biases are $1/2$, then this is a contradiction to Equation (\ref{equ:prob}) for the input $(1, 1, ..., 1)$.
\end{IEEEproof}

\begin{lemma}
\label{lemma6}
Let $P^1$ and $P^2$ be extremal $m$ and $k$-partite full-correlation boxes with associated functions $f_1$ and $f_2$, where $f_1$ depends on the input variables $x_1, x_2, ..., x_m$ and $f_2$ depends on $x_l, x_{l+1}, ..., x_{l+k-1}$ ($l\leq m$). Then the box $P$ with associated function 
\begin{equation}
f(x_1, ..., x_{l+k-1}) = f_1(x_1, ..., x_m)\oplus f_2(x_l, ..., x_{l+k-1})
\end{equation}
is also extremal.
\end{lemma}

\begin{IEEEproof}
We assume that the box with associated function $f$ can be written as a convex combination of two other non-signaling boxes $P_1$ and $P_2$
\begin{equation}
P = \epsilon P_1 + (1 - \epsilon) P_2,
\end{equation} 
where $0 < \epsilon <1$. As before, it is obvious that both of the boxes must fulfill that the XOR of their output elements is equal to the XOR of the Boolean functions $f_1$ and $f_2$. Therefore, we define $f(X_1, ..., X_{l+k+1}) = f_1(X_1,...,X_m) \oplus f_2(X_l, ..., X_{l+k+1})$. We have
\begin{equation}
\text{Prob}\left[ \bigoplus\limits_{i=1}^{n} A_i =f(X_1, ..., X_{l+k+1})\mid X_i = x_i ~\forall ~i \right] = 1
\end{equation}
for all input elements $x_i \in \{0, 1\}$.

Assume that all parties $i \in \{x_m, ..., x_{l+k-1}\}$ input 0 to the box. Therefore, the box acts like the box $P^1$ (assume that the parties $m$ to $k+l-1$ are the same or are able to communicate to each other), and we have found a convex combination of this box. This is in contradiction to the assumption that $P^1$ is extremal. Therefore, the new box is also extremal.
\end{IEEEproof}

Note that the $n$-partite full-correlation box associated to the function $f(x_1, ..., x_n) =\prod_{i=1}^{n}x_i \oplus x_1$ is also an extremal box since it can be constructed with an $n$-PR and an $(n-1)$-PR box by flipping the input bit $x_1$.

\section{Distillation of Full-Correlation Boxes}
\label{sec:dist}
We introduce a new noncommunicative protocol for distillation which requires the parties to arbitrarily distribute the input-and output-interfaces of the weak boxes between the parties. Therefore, the parties have no longer a fixed access to the box, it is even possible that one party has no access to a box, but another one has multiple ones.
\subsection{Distilling $n$-Partite PR Boxes}
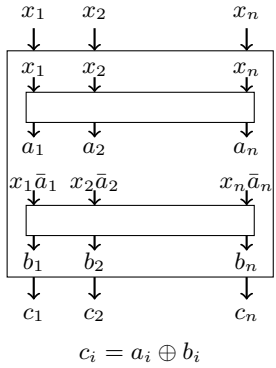
\begin{figure}[!htb]
\begin{center}
\begin{tikzpicture}
\begin{small}
\node [draw, rectangle, minimum width=3.5cm, minimum height=3cm] (A) at (0,0) {};
\node [draw, rectangle, minimum width=3cm, minimum height=0.4cm] (B)  at (0,0.75) {};
\node [draw, rectangle, minimum width=3cm, minimum height=0.4cm] (C) at (0,-0.75) {};
\draw[->, thick] (-1.4,1.8) to[] (-1.4,1.5);
\draw[->, thick] (-0.6,1.8) to[] (-0.6,1.5);
\draw[->, thick] (1.4,1.8) to[] (1.4,1.5);
\node [] () at (-1.4,2) {$x_1$};
\node [] () at (-0.6,2) {$x_2$};
\node [] () at (1.4,2) {$x_n$};

\draw[->, thick] (-1.4,1.15) to[] (-1.4,0.95);
\draw[->, thick] (-0.6,1.15) to[] (-0.6,0.95);
\draw[->, thick] (1.4,1.15) to[] (1.4,0.95);
\node [] () at (-1.4,1.25) {$x_1$};
\node [] () at (-0.6,1.25) {$x_2$};
\node [] () at (1.4,1.25) {$x_n$};

\draw[->, thick] (-1.4,0.55) to[] (-1.4,0.35);
\draw[->, thick] (-0.6,0.55) to[] (-0.6,0.35);
\draw[->, thick] (1.4,0.55) to[] (1.4,0.35);
\node [] () at (-1.4,0.2) {$a_1$};
\node [] () at (-0.6,0.2) {$a_2$};
\node [] () at (1.4,0.2) {$a_n$};

\draw[->, thick] (-1.4,-0.35) to[] (-1.4,-0.55);
\draw[->, thick] (-0.6,-0.35) to[] (-0.6,-0.55);
\draw[->, thick] (1.4,-0.35) to[] (1.4,-0.55);
\node [] () at (-1.4,-0.25) {$x_1\bar{a}_1$};
\node [] () at (-0.6,-0.25) {$x_2\bar{a}_2$};
\node [] () at (1.4,-0.25) {$x_n\bar{a}_n$};

\draw[->, thick] (-1.4,-0.95) to[] (-1.4,-1.15);
\draw[->, thick] (-0.6,-0.95) to[] (-0.6,-1.15);
\draw[->, thick] (1.4,-0.95) to[] (1.4,-1.15);
\node [] () at (-1.4,-1.3) {$b_1$};
\node [] () at (-0.6,-1.3) {$b_2$};
\node [] () at (1.4,-1.3) {$b_n$};

\draw[->, thick] (-1.4,-1.5) to[] (-1.4,-1.8);
\draw[->, thick] (-0.6,-1.5) to[] (-0.6,-1.8);
\draw[->, thick] (1.4,-1.5) to[] (1.4,-1.8);
\node [] () at (-1.4,-2) {$c_1$};
\node [] () at (-0.6,-2) {$c_2$};
\node [] () at (1.4,-2) {$c_n$};
\node [] () at (0,-2.5) {$c_i = a_i \oplus b_i$};

\end{small}
\end{tikzpicture}
\end{center}
\caption[$n$-PR Box distillation]{Generalized BS Protocol for $n$-PR boxes}
\label{fig:dbsprot}
\end{figure}

Using the generalization of the Brunner-Skrzypczyk protocol \cite{BS09} that were presented in \cite{EW13a,EW13b} we are able to distill imperfect $n$-partite PR boxes $P^{\text{PR}}_{n,\varepsilon}$, where
\begin{equation}
P^{\text{PR}}_{n,\varepsilon} = \varepsilon P^{\text{PR}}_{n} + (1 - \varepsilon) P^{\text{c}}_{n}\ ,
\end{equation}
and
\begin{equation}
P^{\text{c}}_n(\textit{\textbf{a}}\vert \textit{\textbf{x}}) = \begin{cases} \frac{1}{2^{n-1}}&\bigoplus\limits_{i}a_i = 0\\0&\text{otherwise.}\end{cases}
\end{equation}
\begin{protocol}[Gen. BS Protocol for $n$-PR Boxes \cite{EW13a,EW13b}]
All n parties share two boxes, where we denote by $x_i$ the value that the $i$th party inputs to the first box and by $y_i$ the value that the $i$th party inputs to the second box. The output bit of the first box for the $i$th party is $a_i$, and the output bit of the second box is $b_i$. The n parties proceed as follows: $y_i =  x_i\bar{a}_i$ and they output, finally, $c_i = a_i \oplus b_i$ (see also Fig.~\ref{fig:dbsprot}).
\label{prot}
\end{protocol}

\begin{theorem}
\label{thm:bsprot}
The generalized BS protocol takes two copies of an arbitrary box $P^{PR}_{n,\varepsilon}$ with $0<\varepsilon<1$ to an $n$-partite correlated non-local box $P^{PR}_{n,\varepsilon'}$ with $\varepsilon'>\varepsilon$, i.e., is distilling non-locality. In the asymptotic case of many copies, any $P^{PR}_{n,\varepsilon}$ with $0<\varepsilon$ is distilled arbitrarily closely to the n-PR box.
\end{theorem}

\subsection{Equivalence Between $n$-and $(n,k)$-PR Boxes}
We show that $k$-PR boxes and $(n,k)$-PR boxes are equi\-valent in the sense that using one of the boxes and shared randomness, the other box can be simulated and \emph{vice versa}. This property is transitive.

Assume $k$ parties share a $k$-PR box, where each of the parties has an input and an output. All of these $k$ parties share a random variable with $n-k$ additional parties. This random variable helps to create a local distribution between the parties that has the property that the XOR of all $n$ outputs is zero. Combining these two distributions with an XOR, the new distribution correspond the distribution of the $(n,k)$-PR box.

To see the opposite implication, let $k$ parties share a $(n,k)$~-~PR box, where every party has an input interface that has influence on the distribution in the end, and the corresponding output. The left inputs (and the corresponding outputs) can be distributed to arbitrary parties. In the end, these parties have to take the XOR of their output to get their final output. Therefore, the $k$ parties simulate a $k$-PR box. 

\subsection{Distillation of $(n,k)$-PR Boxes}
Using the generalized BS Protocol we are able to distill imperfect $(n,k)$-PR boxes $P^{\text{PR}}_{(n,k),\varepsilon}$, where
\begin{equation}
P^{\text{PR}}_{(n,k),\varepsilon} = \varepsilon P^{\text{PR}}_{(n,k)} + (1 - \varepsilon) P^{\text{c}}_{n}\ .
\end{equation}
With the same construction as we showed the equivalence between $n$-and $(n,k)$-PR boxes, we are able to show equivalence between $P^{\text{PR}}_{(n,k),\varepsilon}$ and $P^{\text{PR}}_{k,\varepsilon}$. Because of Theorem \ref{thm:bsprot}, we know that $P^{\text{PR}}_{k,\varepsilon}$ can be distilled arbitrarily closely to $P^{\text{PR}}_{k}$. Again, $P^{\text{PR}}_{k}$ is equivalent to $P^{\text{PR}}_{(n,k)}$. Therefore, $P^{\text{PR}}_{(n,k),\varepsilon}$ can be distilled arbitrarily closely to $P^{\text{PR}}_{(n,k)}$.

\subsection{Distillation of Full-Correlation Boxes}
As shown in Lemma \ref{lem:boolf}, every (imperfect) $n$-partite full-correlation box can be simulated by (imperfect) $n$-PR boxes, where some parties input the constant one. These boxes are exactly the $(n,k)$-PR boxes. If we could isolate each of these (imperfect) $(n,k)$-PR boxes, then we could distill each of them to almost perfect, and so, the whole full-correlation box can be distilled to almost perfect.

Let us assume that $P$ is the full-correlation box, $P^*$ the closest local full-correlation box, and $\varepsilon$ a parameter between 0 and 1. We show that every convex combination
\begin{equation}
P_\varepsilon = \varepsilon P + (1-\varepsilon ) P^*
\end{equation}
can be distilled arbitrarily closely to the full-correlation box $P$.
Without loss of generality, assume that $P$ has no local part, since we change only the local part of the box that can be reached by a local strategy. That implies that $P^* = P_n^{\text{c}}$.

An $(n,k)$-PR box belonging to $a_I$, $I \in \mathcal{J}$, can be isolated from the full-correlation box, if there exist no $J \in \mathcal{J}$ such that $J \subseteq I$. Then, the box can be isolated when every party $i \notin \mathcal{J}$ inputs 0 to the full-correlation box. Otherwise, the box cannot be isolated in this way, but there exist two other possibilities:
Assume there is an $(n,k)$-PR box belonging to $a_I$ and an $(n,l)$-PR box ($k>l$) belonging to $a_J$, $J\subset I$.
\begin{enumerate}
	\item If there exists an $(n,m)$-PR box with $m\geq k$ that can be isolated, then we replace our box with it.
	\item Otherwise, we take the $(n,m)$-PR box with the biggest $m$, isolate or distill a PR-box, and use the recursive construction for $k$-PR boxes \ref{sec:constr}.
\end{enumerate}
Therefore, every imperfect full-correlation box can be distilled in this way and we are also in the multipartite case able to distill boxes that are close to the local bound to maximal non-local ones.

\section{Example}
\label{sec:ex}
\begin{figure}[!htb]
\begin{center}
\begin{tikzpicture}
\begin{small}
\node [draw, rectangle, minimum width=2cm, minimum height=1cm] (A) at (-1.5,0) {$x_1x_2x_3$};
\node [draw, rectangle, minimum width=2.6cm, minimum height=1cm] (E) at (-5.8,0) {$P^1$};
\draw[->, thick] (-3,0) to[] (-4,0);
\draw[->, thick] (-4,0) to[] (-3,0);

\draw[red] (-5.2,0) ellipse (0.6cm and 1.6cm);
\node [red] () at (-6.1,1.5) {3\emph{rd} party};

\draw[->, thick] (-5.5,0.75) to[] (-5.5,0.5);
\draw[->, thick] (-6.1,0.75) to[] (-6.1,0.5);
\draw[->, thick] (-4.9,0.75) to[] (-4.9,0.5);
\draw[->, thick] (-6.7,0.75) to[] (-6.7,0.5);
\node [] () at (-6.7,1) {$x_1$};
\node [] () at (-5.5,1) {$x_3$};
\node [] () at (-6.1,1) {$x_2$};
\node [] () at (-4.9,1) {$0$};

\draw[->, thick] (-5.5,-0.5) to[] (-5.5,-0.75);
\draw[->, thick] (-6.1,-0.5) to[] (-6.1,-0.75);
\draw[->, thick] (-4.9,-0.5) to[] (-4.9,-0.75);
\draw[->, thick] (-6.7,-0.5) to[] (-6.7,-0.75);
\node [] () at (-5.5,-1) {$a_3$};
\node [] () at (-6.1,-1) {$a_2$};
\node [] () at (-4.9,-1) {$a_4$};
\node [] () at (-6.7,-1) {$a_1$};

\draw[->, thick] (-2.3,0.75) to[] (-2.3,0.5);
\draw[->, thick] (-1.5,0.75) to[] (-1.5,0.5);
\draw[->, thick] (-0.7,0.75) to[] (-0.7,0.5);
\node [] () at (-2.3,1) {$x_1$};
\node [] () at (-1.5,1) {$x_2$};
\node [] () at (-0.7,1) {$x_3$};

\draw[->, thick] (-2.3,-0.5) to[] (-2.3,-0.75);
\draw[->, thick] (-1.5,-0.5) to[] (-1.5,-0.75);
\draw[->, thick] (-0.7,-0.5) to[] (-0.7,-0.75);
\node [] () at (-2.3,-1) {$a_1$};
\node [] () at (-1.5,-1) {$a_2$};
\node [] () at (-0.7,-1) {$a_3\oplus a_4$};

\end{small}
\end{tikzpicture}
\end{center}
\caption[Isolation of a 3-PR box]{Isolation of a 3-PR box}
\label{fig:isolation}
\end{figure}
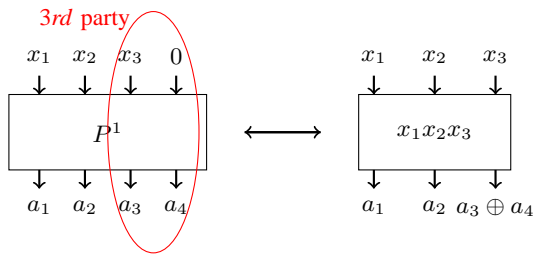

In this example we distill the following full-correlation box:
\begin{equation}
P^1(\textit{\textbf{a}}\vert \textit{\textbf{x}}) = \begin{cases} \frac{1}{8}&\sum\limits_{i=1}^4 a_i = x_1x_2x_3x_4\oplus x_1x_2x_3\oplus x_3x_4\\0&\text{otherwise.}\end{cases}
\end{equation}
The closest local box is $P_4^{\text{c}}$. Let us assume that we have imperfect boxes that are close to the local bound
\begin{equation}
P^1_{\varepsilon} = \varepsilon P^1 + (1-\varepsilon) P_n^{\text{c}} \ .
\end{equation}
First, we isolate the tripartite PR box between the first three parties. Therefore, the first two parties take the first two inputs and the corresponding outputs of the box, and the third party takes the third and fourth inputs and outputs of the box. Into the fourth input, the third party inputs 0 (see Fig. \ref{fig:isolation}). We apply the same method to isolate the bipartite PR box. As soon as the boxes are isolated, they can be distilled to almost  perfect.

Since the 4-PR box cannot be isolated and there does also not exist an $n$-PR box with $n\geq 4$, we have to construct it by smaller ones. Therefore, we use the almost-perfect 3-PR and 2-PR box and use the construction given in Section \ref{sec:constr}. We can now put the three boxes together and receive the almost perfect full-correlation box $P^1$.

\section{Conclusion}
We have studied the problem of non-locality distillation in the
multipartite setting, where the parties are not allowed to communicate
to each other. 
First, we characterized 
maximally non-local
 full-correlation boxes and showed that for every full-correlation box, 
there exists a closest local box which is also a full-correlation
box. 
Second, based on the generalized Brunner/Skrzypzyk protocol, we 
showed that every full-correlation box can be distilled to almost 
perfect without communication if the interfaces of the box can be 
arbitrarily distributed by the parties. This implies that it is 
possible in the multipartite case to distill boxes that are arbitrarily close to
the 
local bound to boxes that are maximally non-local. It remains an 
open question to classify and find distillation protocols for
multipartite 
non-local boxes that are not full-correlation boxes.

\section*{Acknowledgment}
The authors thank Jibran Rashid, Benno Salwey, and Marcel Pfaffhauser for helpful discussions.
This work was supported by the Swiss National Science Foundation (SNF),
the NCCR ``Quantum Science and Technology" (QSIT), and the COST
action on ``Fundamental Problems in Quantum Physics."

\bibliographystyle{IEEEtran}
\bibliography{refs}

\end{document}